\documentclass[11pt, a4paper]{article}
\usepackage{array}
\usepackage{graphicx}
\usepackage{amsmath, amssymb, amsthm}
\usepackage{hyperref}
\usepackage{amsfonts}
\usepackage{amssymb}
\usepackage{amsthm}
\usepackage{amsmath}
\usepackage{multirow}
\usepackage{color}
\usepackage{cite}
\usepackage{float}
\usepackage{geometry}
\usepackage{authblk}
\usepackage{hyperref}
\usepackage{cleveref}
\usepackage{amsmath}   
\usepackage{amssymb}   
\usepackage{siunitx}   

\newcommand{\beq}{\begin{equation}}
	\newcommand{\eeq}{\end{equation}}
\newcommand{\bea}{\begin{eqnarray}}
	\newcommand{\eea}{\end{eqnarray}}

\newcommand{\gsim}{\lower.7ex\hbox{$\;\stackrel{\textstyle>}{\sim}\;$}}
\newcommand{\lsim}{\lower.7ex\hbox{$\;\stackrel{\textstyle<}{\sim}\;$}}
\newcommand{\be}{\begin{equation}}
	\newcommand{\ee}{\end{equation}}
\newcommand{\ba}{\begin{eqnarray}}
	\newcommand{\ea}{\end{eqnarray}}

\usepackage{caption}

\usepackage{cancel}

\title{Micron-Sized Extra Dimensions and Primordial Black Holes: Charged, Rotating, and Memory-Burdened}
\author{George K. Leontaris\thanks{Email: \texttt{leonta@uoi.gr}}, \;George Prampromis\thanks{Email: \texttt{g.prampromis@uoi.gr}}\;}

\affil{${}^*$Physics Department, University of Ioannina, 45110, Ioannina, Greece}

\begin{document}
	
	\maketitle

	\begin{abstract}
		
We explore the possibility of explaining dark matter through six-dimensional (6D) black holes in a theory with two dark dimensions.  Interestingly, in this scenario the fundamental energy scale is of the order of  $M_* \sim 10$ TeV, accessible by future experiments. We analyze the viability of charged and rotating 6D PBHs under the standard Hawking evaporation as well as the memory burden  scenario.	In the case of pure Hawking evaporation, only PBHs with masses $M \gtrsim 10^8$ g survive to the present, while near-extremal configurations extend the lifetime by a factor of $\beta^{-1/2}$, where the parameter $\beta$  characterizes the small deviation from extremality. In  the memory burden scenario evaporation is drastically suppressesed, and sub-gram mass black holes can survive to this day.  At future colliders such as the Future Circular Collider  (FCC), these micro black holes produce characteristic high-multiplicity events with thermal spectra, producing high multiplicity events $\langle N \rangle \sim 45$, allowing direct measurement of $  M_*  $ and the number of extra dimensions. Our results demonstrate that the memory burden mechanism opens up a broad new mass window for light  PBH dark matter. Furthermore, we observe that the mass splitting $  \Delta m $ between Kaluza–Klein modes coincides with the atmospheric neutrino mass scale. Thus, the present scenario provides a unified framework that connects the  Swampland program, early-universe cosmology,  collider physics and low energy phenomenology.

	\end{abstract}

	\newpage

	\section{Introduction}
	The huge discrepancy between the electroweak scale $M_{EW}$ and the four-dimensional Planck scale $M_{Pl}$ -known as the hierarchy problem- remains one of the most significant challenges in theoretical physics. Traditional approaches to this problem often invoke grand unified field theories~\cite{Langacker:1980js}, as well as supersymmetry and supergravity~\cite{Nilles:1983ge}. However, an alternative and compelling framework suggests that the observed weakness of gravity is not fundamental but rather an emergent phenomenon arising from the existence of large extra dimensions~\cite{Antoniadis:1990ew,ADD1,ADD2}. In this context, Standard Model (SM) fields are localized on a three-dimensional brane, while gravity alone propagates into a higher-dimensional bulk. The volume of the extra dimensions dilutes the gravitational force, effectively lowering the fundamental scale of quantum gravity to the TeV range.

    In reference~\cite{Vafa}  a fundamentally different perspective was proposed, which laid the foundations of the Swampland program~\cite{Vafa,Palti,DarkDim}. The latter is a systematic effort to distinguish effective field theories (EFTs) that admit a consistent ultraviolet (UV) completion in quantum gravity from those that can not be embedded in such a framework. The last few years the Swampland program has provided unexpected connections between seemingly unrelated low-energy observables. Amongst the most fascinating is the Dark Dimension scenario~\cite{DarkDim}, which emerges from the combination of the AdS Distance Conjecture and the measured value of the cosmological constant $\Lambda $. This scenario predicts that the smallness of dark energy entails the existence of one extra spatial dimension with a size determined by $ \Lambda $
	\[
	R_{\text{KK}} \sim \Lambda^{-1/4} \sim \mathcal{O}(\mu\text{m}),
	\]
	a scale that is experimentally accessible to tests of Newton's law at sub-millimeter distances. According to the scenario the dark dimension is the largest of the extra dimensions while the remaining ones are assumed to be significantly smaller.

    \noindent 
	A key ingredient in the Swampland program is the Weak Gravity Conjecture (WGC)~\cite{WGC}, which asserts that in any consistent theory of quantum gravity, there must exist a particle with mass $ m \lesssim q M_{Pl} $, where $ q $ is its gauge charge. This conjecture, originally formulated for $ U(1) $ gauge fields, has been generalized to constrain the spectrum of light states in any EFT coupled to gravity. Specifically, the species scale $ \Lambda_{\text{sp}} $ -the cutoff of the EFT where gravitational interactions become strong- is tied to the number of light degrees of freedom. In the presence of a large extra dimension, the tower of Kaluza-Klein (KK) states lowers the species scale relative to the higher-dimensional Planck mass.

	Recently, a specific extension of the original Dark Dimension scenario has been proposed~\cite{Anchordoqui:2025nmb,TwoDarkDim}, consisting of a compact space with two dark dimensions ($\mathfrak{n} = 2 $) of micron scale. Although the relation $R_{\text{KK}} \sim \Lambda^{-1/4} $ is independent of the number of extra dimensions, the case $ \mathfrak{n} = 2 $ yields qualitatively different phenomenology. Most notably, the species scale is reduced to approximately $ \Lambda_{\text{sp}} \sim 10 $ TeV. Remarkably, this value lies  close to an energy scale that, while partially accessible to the current Large Hadron Collider (LHC), is certainly within reach of near-future experiments. It should be mentioned, however, that current null results from the ATLAS and CMS collaborations for mono-jet and mono-photon searches already place a lower bound on this species scale at roughly 4.5 TeV~\cite{ATLAS}. Yet the 10 TeV scale remains partially accessible -particularly through the LHC's high-energy runs (up to 13.6 TeV center-of-mass) and its high-luminosity upgrades- leaving a window of opportunity for discovery.
	In any case, this  reduced species scale  remains a primary target for the proposed FCC~\cite{FCC:2025lpp}. 
	Thus, this two micron-sized extra dimensions scenario is particularly attractive because it may simultaneously address both the gauge and cosmological hierarchy problems.
	
	We should point our however, that beyond collider physics, the micron-scale scenario must navigate stringent astrophysical and cosmological bounds. While traditional observations of Supernova 1987A \cite{Cullen:1999hc} and neutron star heating provide restrictive limits on extra dimensions~\cite{Hannestad:2001xi} (see also~\cite{Hardy:2025ajb}), as suggested in~\cite{DarkDim}, these can be reconciled if the extra dimensions lack isometries~\cite{TwoDarkDim}. This can occur in particular for geometries~\cite{Kachru:2002sk}  beyond the Calabi-Yau (CY) manifolds of the compact space, as is the case of the twisted torii  and the presence of non-geometric fluxes (see eg~\cite{Blumenhagen:2013hva} and ~\cite{AbdusSalam:2024arh}). In such cases,  the  isometries  are not preserved  allowing massive KK gravitons to decay into lighter modes within the  ``dark tower'' itself, effectively evading radiative decay constraints.
	
	Furthermore, this scenario provides a robust landscape for dark matter candidates. In the paper~\cite{Anchordoqui:2025nmb} the authors examine the possibility that the entirety of cosmological dark matter is composed of six-dimensional primordial black holes (PBHs) with masses ranging from $10^{8}$ to $10^{21}$ grams. These higher-dimensional black holes are significantly colder and live longer than their 4D counterparts, offering a unique window into the early universe's evolution. In this paper we will extend this case to 6D charged and rotational counterparts. We will study how these two extra non-compact micron-sized dimensions affect the lifetime of these black holes and analyze the impact of these higher-dimensional objects in assessing the fraction of dark matter that could be composed of PBHs.
	
	Finally, because the quantum gravity scale in this scenario is approximately $\sim 10$ TeV we suppose that micro black holes can be created at accelerators with energy beams approximately $\sim 10$ TeV. We will study the footprints of these micro black holes at future colliders.
	
	The layout of the paper is organized as follows: In section 2 we review the key concepts of the two micro-size extra dimensions scenario,  In section 3, we 
	recapitulate the basics of the Reissner–Nordstr\"om  metric in higher-dimensional space. We 
	review the Schwinger pair production from near-extremal black holes in five dimensions and work out the six-dimensional case exploring in particular their lifetime, firstly for the case of Hawking evaporation and secondly for the memory burden scenario. Section 4 deals with the same issues for the case of rotating black holes. In section 5 we discuss possible  experimental implications of our findings. Section   6 summarizes the results and conclusions.

	\section{Two Micron-Sized Dark Dimensions}
	
	The Dark Dimension scenario emerges from the `interrelation' between the Swampland program and the observed value of the cosmological constant. In this section, we review the key concepts that underpin the extension in two dark dimensions, namely the species scale and the AdS Distance Conjecture.

	In theories with extra dimensions,  the fundamental scale of quantum gravity is controlled by the volume of the compact extra dimensions. More precisely, the fundamental $(4+\mathfrak{n})$-dimensional Planck scale $M_*$, which we will often refer to as species scale  $\Lambda_{\text{sp}}$,  is related to the four-dimensional  $M_{Pl}$ and the volume $V_{\mathfrak{n}}$ of the extra dimensions by
	\begin{equation}\label{species_scale}
		M_* = \Lambda_{\text{sp}} = M_{Pl}^{\frac{2}{2+\mathfrak{n}}} V_n^{-\frac{1}{2+\mathfrak{n}}}.
	\end{equation}
	For a toroidal compactification with equal radii $R$, the volume is $V_{\mathfrak{n}} = (2\pi R)^{\mathfrak{n}}$, yielding,
	\begin{equation}
		M_*^{\mathfrak{n}+2} = \frac{M_{Pl}^2}{(2\pi R)^{\mathfrak{n}}}.\label{Mn}
	\end{equation}
	Thus, from the above formulae we deduce that  large extra dimensions imply  a large compactification volume $V_{\mathfrak{n}}$ which 
	 lowers the fundamental scale $M_*$ relative to $M_{Pl}$.
	
	The AdS Distance Conjecture~\cite{AdSDistance} asserts that in any consistent theory of quantum gravity, as one approaches the boundary of the moduli space (e.g., as the cosmological constant $\Lambda$ tends to zero from negative values), an infinite tower of states becomes light. The mass scale $m_{\text{KK}}$ of this tower is as follows
	\begin{equation}\label{ads_distance}
		m_{\text{KK}} \sim |\Lambda|^{\alpha},
	\end{equation}
	where the exponent $\alpha$ is $\alpha=\frac{1}{4}$~\cite{DarkDim}.   For de Sitter space  a similar relation is expected to hold. By combining this conjecture with the measured value of the cosmological constant $\Lambda$,  one obtains
	\begin{equation}
		m_{\text{KK}} \sim \Lambda^{1/4} \sim 10^{-3}~\text{eV},
	\end{equation}
	which corresponds to a compactification radius $R \sim 1/m_{\text{KK}} \sim 10^{-4}~\text{m} = 100~\mu\text{m}$. However, more refined arguments discussed in~\cite{DarkDim} yield:
	\begin{equation}
		R_{\text{KK}} \sim \Lambda^{-1/4} \sim \mathcal{O}(\mu\text{m}),
	\end{equation}
	which is the micron scale that characterizes the Dark Dimension.
	
	Remarkably, this relation is independent of the number of extra dimensions
    $\mathfrak{n}$~\cite{TwoDarkDim}, which means that any extra dimension(s) of micron scale must satisfy this bound.
	\\
	
	Besides,  the Weak Gravity Conjecture (WGC)~\cite{WGC} is one of the most significant Swampland conjectures. In fact, by virtue of the WGC  a consistency condition is imposed  on any effective field theory coupled to gravity that admits a UV completion in quantum gravity. In its simplest form, the WGC states that gravity must be the weakest force. More precisely, for a $U(1)$ gauge field, there must exist a particle with mass $m$ and charge $q$ such that:
	\begin{equation}\label{wgc_condition}
		q \ge m~,
	\end{equation}
	(in units where $M_{Pl}=1$, or equivalently, $q \ge  m/M_{Pl}$, in Planck units).
	This condition ensures that extremal black holes, i.e., those with $M = Q$ can decay, thereby avoiding the formation of stable remnants or naked singularities. 
	To further analyse  the physical motivation of the WGC as well as its implications, consider an extremal black hole of mass $M$ and charge $Q$ satisfying $M = Q$ (in units where $4\pi G = 1$). Such a black hole has zero temperature and is classically stable. However, if it cannot decay, it would become a stable remnant which constitutes a potential problem for quantum gravity. The WGC ensures that there exists a charged particle that allows the black hole to shed charge faster than it loses mass. Furthermore,  suppose the black hole emits a particle of mass $m$ and charge $q$, thus,  after emission the BH mass and charge are given by
	\begin{align}
		M' = M - m &;\;
		Q' = Q - q.
	\end{align}
	To avoid violating the extremality bound $M' \ge Q'$  which would lead to a naked singularity, we require
	\begin{equation}
		M - m \ge Q - q.
	\end{equation}
	For an initially extremal black hole, $M = Q$, this simplifies to
	\begin{equation}
		-m \ge -q \quad \Longrightarrow \quad q \ge m.
	\end{equation}
	Thus, for the black hole to be able to decay without violating cosmic censorship, there must exist a particle with charge greater than or equal to its mass. This is precisely the WGC condition.
	
	In the context of the Dark Dimension scenario, the WGC plays a crucial role in constraining the spectrum of light states. Specifically, the presence of a single large extra dimension implies a tower of KK modes. The WGC requires that these modes satisfy certain charge-to-mass ratios, which in turn bound the scale $M_*$ and the number of extra dimensions.
	
	In the scenario with two  micron-sized extra dimensions ($\mathfrak{n}=2$) and fundamental scale $M_* \sim 10$ TeV, the WGC
    requires the existence of a state with charge $q$ and mass $m$  satisfying $q \ge m/M_*$. 
  This condition can be  naturally satisfied by  KK modes, whose masses are of order $\sim 1/R \sim 10^{-3}$ eV.
  If the relevant $U(1)$ gauge field arises from compactification, or if charged bulk fields are present,
  these KK modes carry charge and provide the required super-extremal states.

	Returning now to  the specific extension proposed in reference~\cite{TwoDarkDim}, the authors consider $\mathfrak{n}=2$ extra dimensions of micron scale. The AdS Distance Conjecture fixes the compactification radius to $R \sim \Lambda^{-1/4} \sim 10^{-4}~\text{cm}$, independently of $\mathfrak{n}$. This yields:
	\begin{equation}
		M_* = \left( \frac{M_{Pl}^2}{R^2} \right)^{1/4}  \sim 10~\text{TeV}.
	\end{equation}
	Remarkably, now the fundamental scale of quantum gravity is lowered down to the $M_*\sim 10$ TeV range, a scale that solves the hierarchy problem and at the same time is phenomenologically accessible to current and future colliders.
	
	Therefore, from the above analysis, we observe that in the scenario of $\mathfrak{n}$ extra micron-sized dimensions,  the combination of the AdS Distance Conjecture and the Weak Gravity Conjecture provides a   consistent  framework with  characteristic scales.  In particular, the  size of the extra dimensions is tied to the observed dark energy scale, while  the fundamental Planck scale is reduced to $M_*\sim 10$ TeV, which does not enormously differ from the electroweak scale. Moreover, there exists a tower of KK states that satisfy the WGC, while -as we will also confirm in the subsequent analysis- primordial black holes in six dimensions can constitute all of the dark matter. 
	Thus,  provided the extra dimensions lack continuous isometries, the scenario is cosmologically viable. 
	This theoretical consistency, combined with the rich phenomenology to be discussed in the following sections, makes the $\mathfrak{n}=2$  scenario a compelling framework for physics beyond the Standard Model.
	
	Closing this section, for later convenience, we summarize  the  key scales of the single and two micron-sized scenarios in Table~\ref{tab:scales}. 
	
	\begin{table}[htbp]
		\centering
		\begin{tabular}{ll}
			\textbf{Physical parameter} &  \textbf{Order of magnitude scale} \\
			\hline
			Cosmological Constant: & $\Lambda\,\lesssim\, 10^{-122} M_{Pl}^4$ \\
					Dark Dimension radius: & $R \sim 1~\mu\text{m}$ \\
			KK mass scale $\sim  1/R$: & $m_{\text{KK}} \sim 10^{-3}~\text{eV}$ \\
				5D Planck scale: & $M_* \sim 10^{10}~\text{GeV}$ \\
			6D Planck scale: & $M_* \sim 10~\text{TeV}$ \\
			4D reduced Planck mass:& $\overline{M}_P \sim 2.4\cdot  10^{18}~\text{GeV}$ \\
		\end{tabular}
		\caption{The various scales  in the  scenarios with one and two extra dimensions. The Dark Dimension scenario provides a unified framework that connects these seemingly disparate scales through the Swampland conjectures.}
		\label{tab:scales}
	\end{table}

	\section{Higher-Dimensional Charged Black Holes}
	
	The Reissner–Nordstr\"om (NR) metric is an exact solution to the coupled Einstein–Maxwell equations in general relativity. It describes the spacetime geometry around a non-rotating, spherically symmetric body that has both mass $M$ and electric charge $Q$ (in geometric units where $G = c = 1$), which can be considered as  the charged generalization of the Schwarzschild metric. This metric is static, meaning that there is no time dependence, and asymptotically flat so that far away, spacetime looks like Minkowski space.
    \subsection{Bounds in de Sitter Universe}	
Before proceeding to the analysis of charged black holes in five and six dimensions, it is important to highlight another significant mass bound that arises in de Sitter space from the Festina Lente (FL) conjecture~\cite{Montero:2019ekk}.

In $D$-dimensions, the RN-dS metric is
\begin{align}
    ds^2 = U(r) dt^2 - \frac{dr^2}{U(r)} - r^2 d\Omega^2_{D-2}~,
\end{align}
where $d\Omega^2_{D-2}$ is the line element of a $(D-2)$-dimensional unit sphere and
\begin{align}
    U(r) = 1 - \frac{2M}{M_*^{D-2} r^{D-3}} + \frac{(e' Q)^2}{4\pi M_{*}^{D-2} r^{2D-6}}-\frac{r^2}{\ell_D^2}\label{metricRN}.
\end{align}
where in the last term $\ell_D$  is the dS radius.
\paragraph{}Here, $M$ is the black hole mass, $Q$ its electric charge, $e'$ is a dimensionless normalization factor absorbing gauge-coupling conventions, $M_*$ is the $D$-dimensional reduced Planck mass, and $\ell_D$ is the de Sitter radius. The parameter $m_{e'}$  that appears in the Festina Lente bound below is the physical mass of the lightest charged particle in the theory, which couples to the gauge field with strength $e'$; it is not a parameter of the classical metric but rather a dynamical field-theory mass that enters through the Schwinger pair-production rate.

When a black hole is embedded in a de Sitter (dS) background, quantum gravity imposes a fundamental constraint~\cite{Montero:2019ekk} which translates into a lower bound on the mass parameter
$m_{e'}$.
 This conjecture constitutes a central element of the Swampland program for theories with dS vacua and is closely tied to the behavior of charged black holes.

A key ingredient in the derivation of the FL bound is the Nariai black hole~\cite{Nariai:1950}. A Reissner--Nordstr\"om--de Sitter (RN--dS) black hole possesses both a black hole event horizon and a cosmological horizon. In the Nariai limit, these two horizons coincide, corresponding to the largest black hole that can exist in a de Sitter universe.
As argued in~\cite{Montero:2019ekk}, requiring that this extremal configuration can decay via Hawking radiation emitted by the cosmological horizon, without becoming super-extremal and destabilizing the spacetime, leads directly to the FL bound.
Physically, the Festina Lente conjecture implies that charged particles in a de Sitter background cannot be arbitrarily light. Since there exists a maximum black hole size set by the Nariai limit~\cite{Nariai:1950}, sufficiently large charge-to-mass ratios could otherwise drive the black hole beyond extremality. To prevent this, black holes must be able to discharge efficiently through Schwinger pair production. Requiring that this discharge process occurs faster than the competing mechanisms that increase the black hole's charge leads to a bound on the mass of any charged particle in a theory with a de Sitter background. Parametrically, this condition takes the form

\begin{equation}
m^4_{e'}\, \gtrsim\, (e'q)^2 \frac{(D-1)(D-2)}{2 M_P^{2-D}\ell_D^2}\label{FL} 
\end{equation}

A second constraint arises from the WGC which requires the existence of a charged particle with charge-to-mass ratio at least that of an extremal black hole. In $D=(4+\mathfrak{n})$-dimensions, this yields: 
\begin{equation}
        \frac{g q}{m}\, \gtrsim \,\sqrt{\frac{D-3}{D-2}} \frac{1}{M_P^{\frac{D-2}2}}\label{WGCbound}
\end{equation}
We deduce that equation~(\ref{WGCbound}) imposes an upper bound on the mass of the lightest charged particle, $m \lesssim g q M_P^{(D-2)/2}$. This must hold in any consistent quantum gravity theory, independently of the background curvature.	

It is worth noting that the dimensional scaling of the gauge coupling, $[g] = E^{2-D/2}$, ensures that the FL inequality~(\ref{FL}) remains invariant under changes of spacetime dimension~\cite{MonteroVafa}.
Crucially, however, the FL bound is phenomenologically relevant only for black holes whose radii approach the cosmological horizon.

%
%
This renders the bound largely decoupled from the physics at scales much smaller than $  R_{KK}  $. While the de Sitter Weak Gravity Conjecture (dS-WGC) typically constrains black holes with horizons smaller than $  R_{KK}  $~\cite{Antoniadisbenakli}, its impact on the particle spectrum becomes negligible in the limit of large $  \ell_4  $.
Consequently, the subsequent analysis assumes a nearly flat five-dimensional Minkowski background, which allows us to omit the final term in Eq.~(\ref{metricRN}).

	\subsection{Higher-Dimensional  Schwinger Effect and Black Hole Thermodynamics for Higher-Dimensional Charged Black Holes }
	
	To facilitate our subsequent analysis, first we proceed with the generalization of some basic concepts in higher dimensional spacetime, starting with the Schwinger Effect and Black Hole Thermodynamics.
	
	The classic Schwinger formula~\cite{Schwinger:1951} in four dimensions gives the rate of pair production in a constant electric field for a particle of spin $J$, mass $m$, and charge $e$. Here we are interested in generalizations of this formula in a higher-dimensional background. This would involve the electric field $E'$ associated with the charge of the Reissner-Nordstr\"om  black hole, which depends on the distance $r$ from the source. A string-theoretic derivation of the formula is given by Bachas and Porrati~\cite{Bachas:1992}
	\begin{equation}\label{Schwinger}
		\Gamma_D = \frac{2j+1}{(2\pi)^{D-1}} \sum_{n=1}^{\infty} (-1)^{(2j+1)(n+1)} \left(\frac{e'E'}{n}\right)^{D/2} \exp\left(-\frac{n\pi m_{e'}^2}{e'E'}\right),
	\end{equation}
	where $D = 4 + \mathfrak{n}$ is the total number of spacetime dimensions, while $\mathfrak{n}$ stands for the extra dimensions. In this context, the factor $(e'E'/n)^{D/2}$ emerges from the $t^{-D/2}$ dependence in the annulus amplitude, which comes from integrating over the $D-2$ transverse momenta and the zero modes. The exponential contains $m_{e'}^2$, reflecting the instanton action, while the spin degeneracy $2j+1$ together with the sign factor $(-1)^{(2j+1)(n+1)}$ account for the spin statistics.

    Next, we mention the generalizations of the fundamental physical quantities that will be used in the analysis that follows.
	
	The horizon radius of a $D$-dimensional Schwarzschild black hole is
	\begin{equation}\label{radius0}
		r_s = \frac{1}{\sqrt{\pi} M_*} \left( \frac{M}{M_*} \right)^{\frac{1}{D-3}} \left( \frac{8\Gamma\left(\frac{D-1}{2}\right)}{D-2} \right)^{\frac{1}{D-3}}.
	\end{equation}
	For our purposes, however, it suffices to simply write
	\begin{equation}
		r_s \sim \frac{1}{M_*} \left(\frac{M}{M_*}\right)^{\frac{1}{D-3}}.\label{radius}
	\end{equation}
Then, the entropy  is given by
	\begin{equation}\label{entropy}
		S_{\text{BH}} = \frac{1}{4G_D} A = \frac{1}{4G_D} \cdot \Omega_{D-2} r_s^{D-2},
	\end{equation}
    In the last step of the above equation, we have substituted the $D$-dimensional Newton's constant 
    with $G_D = 1/(8\pi M_*^{D-2})$ and  the  the area of the unit $(D-2)$-sphere with $\Omega_{D-2} = 2\pi^{(D-1)/2}/\Gamma((D-1)/2)$ . Using the horizon radius from (\ref{radius}), this yields the scaling relation
	\begin{equation}\label{entropy_scaling}
		S_{\text{BH}} = \frac{\Omega_{D-2}}{4G_D M_*^{D-2}} \left(\frac{M}{M_*}\right)^{\frac{D-2}{D-3}} \equiv \kappa_0 \left(\frac{M}{M_*}\right)^{\frac{D-2}{D-3}},
	\end{equation}
	where $\kappa_0$ is a dimensionless constant depending on $D$. 
    In the subsequent analysis, only the parametric dependence of the entropy is relevant. 
    We therefore retain the scaling
\begin{equation}
S_{\text{BH}} \sim \left(\frac{M}{M_*}\right)^{\frac{D-2}{D-3}}~,
\end{equation}
which makes explicit the dependence of the black hole entropy on the dimensionless ratio
 of the black hole mass to the higher-dimensional Planck scale
$M/M_*$.

	The temperature of a static higher-dimensional black hole, expressed in terms of the BH entropy, is
	\begin{equation}\label{temp}
		T_S = \frac{D-3}{4\pi r_s} = \frac{D-3}{4\pi} \sqrt{\pi} M_* \left( \frac{M_*}{M} \right)^{\frac{1}{D-3}} \left( \frac{D-2}{8\Gamma\left(\frac{D-1}{2}\right)} \right)^{\frac{1}{D-3}}.
	\end{equation}
	In terms of entropy, using $S_{\text{BH}} \propto r_s^{D-2}$,
	\begin{equation}
		T_S = \frac{D-3}{4\pi} \left( \frac{4G_D}{\Omega_{D-2}} \right)^{\frac{1}{D-2}} S_{\text{BH}}^{-\frac{1}{D-2}} \equiv \tau_0 \, M_* \, S_{\text{BH}}^{-\frac{1}{D-2}},
	\end{equation}
	where $\tau_0$ is a dimensionless constant. As already pointed out, we are only interested in the parametric dependence, thus the above is simplified as (we also drop index `BH' for notational  simplicity)
	\begin{equation}
		T_S \sim M_* \, S^{-\frac{1}{D-2}}.\label{temp}
	\end{equation}

	The Schwarzschild black hole decay rate (energy emission rate) is approximately \cite{Anchordoqui:2024}
	\begin{equation}\label{decay_rate}
		\Gamma_S \sim T_S,
	\end{equation}
	up to greybody factors and the number of emitted species, which we absorb into an $\mathcal{O}(1)$ constant.

	The Hawking evaporation time characterizing the lifetime of a bulk static primordial black hole
    is estimated by integrating the energy loss equation $dM/dt = -\Gamma_S$. 
    Using $\Gamma_S \sim T_S \sim M_* (M/M_*)^{-1/(D-3)}$, we obtain
	\begin{equation}\label{lifetime}
		\tau_{\text{BH}} \sim r_s S_{\text{BH}} \sim \frac{1}{M_*} \left(\frac{M}{M_*}\right)^{\frac{D-1}{D-3}}~,
	\end{equation}
	where use of (\ref{entropy}) and (\ref{radius}) was made.

	\subsection{Near-Extremal Black Holes}
	
	For a near-extremal Reissner-Nordstr\"om black hole in $D$ dimensions, the temperature scales as
	\begin{equation}\label{temp_ne_general}
		T_{\text{ne}} \sim \frac{c}{S},
	\end{equation}
	where $c$ encodes the deviation from extremality. For a charged black hole with charge $Q$, one has
	\begin{equation}\label{c_parameter}
		c = \sqrt{M^2 - \frac{(e'Q)^2 M_*^{D-2}}{4\pi}}~\cdot 
	\end{equation}
	For the non-extremal case, where $(e'Q) M_*^{(D-2)/2} / (M\sqrt{4\pi}) \ll 1$, we have $c \sim M$. Using the entropy relation (\ref{entropy}), this yields
	\begin{equation}
		c \sim M_* \, S^{\frac{D-3}{D-2}}~.
	\end{equation}

	\paragraph{}For the near-extremal case, which is of our interest, it holds $M \sim (e'Q) M_*^{(D-2)/2} / \sqrt{4\pi}$.
    In $D$ dimensions, from~(\ref{entropy}) we observe that the standard Schwarzschild black hole entropy scales with mass as $S \sim \left(\frac{M}{M_*}\right)^{\frac{D-2}{D-3}}$. Inverting this formula to express the mass as a function of the entropy yields, at leading order

\begin{equation}
		M(S) = M_* S^{\frac{D-3}{D-2}}.\label{MofS}
	\end{equation}
    We consider a tower of states with masses and charges ($m_n$, $q_n$) with $n=1,2.....N_{sp}$.
     For large but finite entropy $S \gg 1$, the total mass and charge of the minimal black black, being the same as the total species
mass and charge, are then given as~\cite{Ibasile}:
\begin{align}
    M &=M_{sp} = \sum_{n=1}^{N_{sp}} m_n = M_*\mathcal{S}_{sp}^{\frac{D-3}{D-2}} + M_*\mathcal{S}_{sp}^{-\frac{1}{D-2}} + \dots  \\ 
    Q_{\text{eff}}&=Q_{sp} = \sum_{n=1}^{N_{sp}} q_n = M_*\mathcal{S}_{sp}^{\frac{D-3}{D-2}} + M_*(1+\beta)\mathcal{S}_{sp}^{-\frac{1}{D-2}} + \dots 
\end{align}

For a Kaluza-Klein-like tower of charged particles with mass spectrum given by $m_n = m_0 n$, assuming that the electric charges of the particles are shifted as $q_n = m_0 (n + \beta)$ \cite{Ibasile}. For near-extremal charged black holes, both the mass $M$ and the effective gauge charge $Q_{\text{eff}}$ share the exact same leading-order scaling. To represent the subtle gap between mass and charge that keeps the black hole near-extremal rather than strictly extremal, a shift parameter $\beta$ is introduced in the sub-leading term:
\begin{align}
    \beta = \frac{q_n - m_n}{m_0}
\end{align}
Then, we have:
\begin{align}
    M_{sp}^2 &= M_*\mathcal{S}_{sp}^{2\frac{D-3}{D-2}} + 2M_* \mathcal{S}_{sp}^{\frac{D-4}{D-2}} + \dots \\ Q_{sp}^2 &= M_*\mathcal{S}_{sp}^{2\frac{D-3}{D-2}} + 2(1+\beta) M_*\mathcal{S}_{sp}^{\frac{D-4}{D-2}} + \dots
\end{align}
Subtracting the above equations, the leading terms cancel out and it turns out that  the parameter $c$ is proportional to the square root of the factor $\beta$ measuring the gap of the BH from the near extremal limit. Thus $c$  is given by:
\begin{align}
    c= \sqrt{\vert{}M_{sp}^2 - Q_{sp}^2\vert{}} \sim \sqrt{\beta}M_*\left( \mathcal{S}_{sp}^{\frac{D-4}{D-2}} \right)^{1/2} = \sqrt{\beta} M_*\, \mathcal{S}_{sp}^{\frac{D-4}{2(D-2)}}
\end{align}
and in the final form:
\begin{align}
    c \sim M_* \sqrt{\beta} \, S^{\frac{D/2 - 2}{D-2}}
\end{align}
	Combining (\ref{temp_ne_general}) with the expression for $c$, we obtain the near-extremal temperature scaling
	\begin{equation}\label{temp_ne}
		T_{\text{ne}} \sim \frac{ \sqrt{\beta} \, M_* \, S^{\frac{D/2-2}{D-2}}}{S} =  \, M_* \, S^{-1/(D-2)}\sqrt{\frac{\beta}{S}}.
	\end{equation}
   
	Comparing with the static black hole temperature (\ref{temp}), we find
	\begin{equation}
		T_{\text{ne}} = T_S \, \sqrt{\frac{\beta}{S}}~\cdot 
	\end{equation}
	Thus, the near-extremal black hole temperature is suppressed relative to the Schwarzschild temperature by a factor $\sqrt{\beta/S}$, which can be very small for large entropy $S$.

	For a near-extremal black hole, the electric field near the horizon is weak. The Schwinger pair production rate (\ref{Schwinger}) can be evaluated using the near-horizon electric field $E' \sim Q / r_s^{D-2}$. The exponential suppression factor $\exp(-\pi m_{e'}^2 / (e'E'))$ becomes significant, and the dominant contribution comes from the $\mathfrak{n}=1$ term. The rate can be expressed in terms of the black hole parameters as
	\begin{equation}
		\Gamma_{\text{Sch}} \sim \left(\frac{e'Q}{r_s^{D-2}}\right)^{D/2} \exp\left(-\frac{\pi m_{e'}^2 r_s^{D-2}}{e'Q}\right).
	\end{equation}
	This expression highlights the connection between black hole geometry and vacuum instability, which may have important implications for black hole evaporation and the production of charged particles in the early universe.

	\subsubsection{5D Near-Extremal Black Holes}
	
	For $D = 5$, the Schwinger pair production rate in a constant electric field becomes
	\begin{equation}\label{schwinger5d}
		\Gamma_5 = \frac{1}{8\pi^4} \sum_{n=1}^{\infty} \left(\frac{e'E'}{n}\right)^{5/2} \exp\left(-\frac{\pi n m_{e'}^2}{e'E'}\right).
	\end{equation}
	
	The outer horizon radius of a 5D Reissner-Nordstr\"om black hole is given by
	\begin{equation}\label{horizon5d}
		r_{+,5} = \left[\frac{M + \sqrt{M^2 - (e'Q)^2 M_*^3 / (4\pi)}}{M_*^3}\right]^{1/2} \xrightarrow[M \gg (e'Q)M_*^{3/2}]{ } \sqrt{\frac{M}{M_*^3}}.
	\end{equation}
	For comparison, the 4D Schwarzschild horizon scales as $r_{+,4} \sim M/M_*^2$. This implies that for a given mass, the outer horizon of a 5D RN black hole is larger than that of a 4D black hole. Consequently, the electric field strength at the horizon is smaller in five dimensions, making Schwinger pair production less efficient and potentially easier to suppress.
	
	In Ref.~\cite{TwoDarkDim}, the authors derived the lifetime of a static 5D Schwarzschild black hole
	\begin{equation}\label{lifetime5d}
		\tau_s \sim 13.8 \left(\frac{M}{10^{12}\,\text{g}}\right) \left(\frac{6}{\sum_i c_i \Gamma_s f}\right)~\text{Gyr},
	\end{equation}
	where $c_i$ are the degrees of freedom for emitted particles, $\Gamma_s$ is the greybody factor, and $f$ is a normalization factor. They further showed that a dark matter interpretation in terms of 5D static black holes requires masses in the range
	\begin{equation}\label{massrange5d}
		10^{14} \lesssim M/\text{g} \lesssim 10^{21}.
	\end{equation}
	
	Below we summarize the results for 5D Near Extremal PBHs case \cite{Anchordoqui:2024}  and we see how near-extremality modifies this mass range. Consider a black hole with mass $M \sim 10^5$ g. From equation~(\ref{temp}), the Schwarzschild temperature is $T_S \sim 4$ GeV. For such a black hole, the degrees of freedom contributing to Hawking radiation include photons (2), neutrinos (6), charged leptons (12), quarks (48), and gluons (24), yielding $\sum_i c_i(T_S) \approx 45$. The lifetime of a $10^5$ g Schwarzschild black hole is $\tau_s \sim 10^{-5}$ yr. For a near-extremal black hole of the same mass, using equation~(\ref{temp_ne}), the temperature becomes
	\begin{equation}
		T_{\text{ne}} \sim 10^{-5} \sqrt{\beta}~\text{eV},
	\end{equation}
	where the parameter $\beta$ characterizing the small deviation from extremality is assumed to be $\beta< 1$. The corresponding lifetime is estimated to be 
	\begin{equation}\label{lifetime5d_ne}
		\tau_{\text{ne}} \sim \frac{15}{\sqrt{\beta}}~\text{Gyr}.
	\end{equation}
	Thus, for near-extremal 5D black holes to constitute all of the dark matter, the viable mass range extends to
	\begin{equation}\label{massrange5d_ne}
		10^5 \sqrt{\beta} \lesssim M/\text{g} \lesssim 10^{21}.
	\end{equation}
	
	\subsubsection{6D Near-Extremal Black Holes}
	
	We now turn to the case of two extra non-compact dimensions ($\mathfrak{n}=2$, $D=6$). As discussed earlier, the higher-dimensional Planck scale in this scenario is
	\begin{equation}\label{planck6d}
		M_* \sim 10~\text{TeV}.
	\end{equation}
	
	The horizon radius for a 6D RN black hole follows from solving $g^{rr}=0$
	\begin{equation}\label{horizon6d}
		U(r) = 1 - \frac{2M}{M_*^{4} r^{3}} + \frac{(e'Q)^2}{4\pi M_*^{4} r^{6}} = 0 \quad \xrightarrow{D=6} \quad r_{+,6} = \left(\frac{M}{M_*^4}\right)^{1/3}.
	\end{equation}
	
	The lifetime of a static 6D Schwarzschild black hole is given in~\cite{TwoDarkDim} as,
	\begin{equation}\label{lifetime6d}
		\tau_s \sim 13.7 \left(\frac{M}{10^8\,\text{g}}\right)^{5/3}~\text{Gyr}.
	\end{equation}
	Consequently, static 6D PBHs can constitute all of the dark matter for masses in the range
	\begin{equation}\label{massrange6d}
		10^8 \lesssim M/\text{g} \lesssim 10^{21}.
	\end{equation}
	
	For near-extremal black holes, we first compute the Schwarzschild temperature from Eq.~(\ref{temp})
	\begin{equation}\label{temp6d}
		T_s \sim M_* S^{-1/4},
	\end{equation}
	and thus, the decay rate is of the order
	\begin{equation}\label{gammas6d}
		\Gamma_s \sim T_s \sim M_* S^{-1/4}.
	\end{equation}
	The entropy for a 6D black hole is (\ref{entropy}):
	\begin{equation}\label{entropy6d}
		S \sim \left(\frac{M}{M_*}\right)^{4/3}.
	\end{equation}
	Hence, for a BH mass, say $M \sim 1$ g, with $M_* \sim 10$ TeV $\sim 10^{13}$ eV $\sim 10^{-20}$ g (in natural units), we obtain $S \sim 10^{27}$.
	
	Using Eq.~(\ref{temp_ne}), the near-extremal temperature is
	\begin{equation}\label{temp6d_ne}
		T_{\text{ne}} \sim \frac{T_s}{\sqrt{S}} \sqrt{\beta}.
	\end{equation}
	The decay rate is then $\Gamma_{\text{ne}} \sim T_{\text{ne}}$. For a near-extremal black hole with initial mass $M \sim 1$ g, we have $T_S \sim 4 \times 10^{-6}$ GeV and $S \sim 10^{27}$, and $T_{ne}\sim 10^{-7}\sqrt{\beta}~$eV. Using Eq.~(\ref{lifetime6d}) and the relation $\tau_{\text{ne}} \sim (T_s/T_{\text{ne}}) \tau_s$ \cite{Anchordoqui:2024} \footnote{We clarify that $\tau_{ne} \sim \tau_S (T_S / T_{ne})$ was employed as a phenomenological order-of-magnitude scaling approximation based on the quantum emission rate per species $\Gamma \sim T$~\cite{Anchordoqui:2024}. }, we obtain
	\begin{equation}\label{lifetime6d_ne}
		\tau_{\text{ne}} \sim \frac{13.7}{\sqrt{\beta}}~\text{Gyr}.
	\end{equation}
	Thus, near-extremal 6D PBHs can constitute all of the dark matter for masses in the range:
	\begin{equation}\label{massrange6d_ne}
		\beta^{3/14} \lesssim M/\text{g} \lesssim 10^{21}.
	\end{equation}
	
	\subsection{Memory Burden Effect for Near-Extremal Black Holes in $D$ Dimensions}
	
	Having calculated the lifetime of near-extremal black holes, we now incorporate the memory burden effect~\cite{Dvali:2018xpy,Dvali:2020wft,Dvali:2024hsb}, which slows down evaporation when the lifetime of black hole is equal to Page time~\cite{Page:1983uc,Page:1993wv}. The quantum decay rate under the memory burden effect is suppressed by a power of the entropy,
	\begin{equation}\label{gamma_ne_mb}
		\Gamma_{\text{ne}}^p = \frac{\sqrt{\beta}}{S^{p+1/2}} \Gamma_s \sim \frac{\sqrt{\beta}}{S^{p+1/2}} T_s,
	\end{equation}
	where $p$ is a model-dependent exponent characterizing the strength of the memory burden effect (where  $p \geq 1$, with $p=1$ being the minimal value for the MB scenario to be effective). From this, the ratio of temperatures is
	\begin{equation}
		\frac{T_s}{T_{\text{ne}}} \sim \frac{S^{p+1/2}}{\sqrt{\beta}}.
	\end{equation}
	Using $\tau_{\text{ne}} \sim (T_s/T_{\text{ne}}) \tau_s$, we obtain the lifetime under the memory burden effect
	\begin{equation}\label{lifetime_ne_mb}
		\tau_{\text{ne}} \sim \frac{S^{p+1/2}}{\sqrt{\beta}} \tau_s.
	\end{equation}
    \begin{figure}[h]
	\centering
	\includegraphics[width=0.8\textwidth]{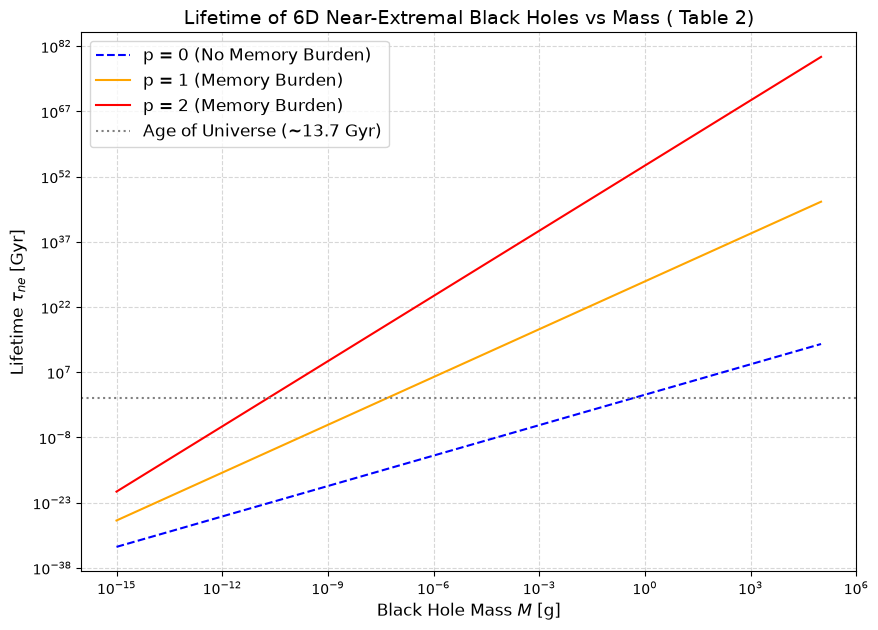}
	\caption{ illustrates the lifetime of 6D rotating primordial black holes across a broad mass spectrum ($10^{-15} \text{ g} \le M_{BH} \le 10^6 \text{ g}$). Under standard Hawking evaporation ($p=0$), near-extremality already suppresses the temperature by $\sqrt{\beta/S}$, shifting the threshold for survival down to $M \gtrsim \beta^{3/14} \text{ g}$. When memory burden is incorporated ($p=1, 2$), the combined suppression factor $\sqrt{\beta}/S^{p+1/2}$ reduces the evaporation rate even further. As shown in the plot, the survival threshold ($\tau_{ne} \ge 13.7 \text{ Gyr}$) drops significantly into the ultra-light sub-gram regime ($M \gtrsim 10^{-7.5} \text{ g}$ for $p=1$ and $M \gtrsim 10^{-10.5} \text{ g}$ for $p=2$. This visual comparison underlines how the memory burden effect opens up a vast new mass window for near-extremal PBH dark matter in two extra dark dimensions.. }
	\label{PBHlifeA}
\end{figure}
	
	For the 6D case, using $\tau_s$ from Eq.~(\ref{lifetime6d}) and $S$ from Eq.~(\ref{entropy6d}), we find
	\begin{align}
		\tau_{\text{ne}} &\sim \frac{1}{\sqrt{\beta} M_*} \left(\frac{M}{M_*}\right)^{(p+1/2)(4/3) + 5/3} \nonumber \\
		&\sim \frac{1}{\sqrt{\beta} M_*} \left(\frac{M}{M_*}\right)^{\frac{4p}{3} + \frac{7}{3}}.
	\end{align}
	For the specific values of $p$ we find:
	
	\begin{itemize}
		\item \textbf{$p=0$ (no memory burden)}; $\tau_{\text{ne}} \sim \frac{1}{\sqrt{\beta} M_*} (M/M_*)^{7/3}$. Taking $M_* = 10$ TeV, this gives:
		\begin{equation}
			\tau_{\text{ne}} \sim \frac{21.3}{\sqrt{\beta}} \left(\frac{M}{\text{g}}\right)^{7/3}~\text{Gyr}.
		\end{equation}
		
		\item \textbf{$p=1$}: $\tau_{\text{ne}} \sim \frac{1}{\sqrt{\beta} M_*} (M/M_*)^{11/3}$. Numerical evaluation yields:
		\begin{equation}
			\tau_{\text{ne}} \sim \frac{2.5 \times 10^{27}}{\sqrt{\beta}} \left(\frac{M}{\text{g}}\right)^{11/3}~\text{Gyr}.
		\end{equation}
		
		\item \textbf{$p=2$}: $\tau_{\text{ne}} \sim \frac{1}{\sqrt{\beta} M_*} (M/M_*)^{5}$. Finally, the numerical value for our last choice $p=2$ is:
		\begin{equation}
			\tau_{\text{ne}} \sim \frac{1.1 \times 10^{54}}{\sqrt{\beta}} \left(\frac{M}{\text{g}}\right)^{5}~\text{Gyr}.
		\end{equation}
	\end{itemize}
	For a visual overview of the PBHs' life spans due to memory burden effect, we summarize the results in  Table~\ref{tab:lifetimes} .
	\begin{table}[htbp]
		\centering
		\begin{tabular}{|c|c|}
			\hline
			\textbf{$p$} & \textbf{$\tau_{\text{ne}}$} \\
			\hline
			0 & $\displaystyle \frac{21.3}{\sqrt{\beta}} \left(\frac{M}{\text{g}}\right)^{7/3}~\text{Gyr}$ \\
			\hline
			1 & $\displaystyle \frac{2.5 \times 10^{27}}{\sqrt{\beta}} \left(\frac{M}{\text{g}}\right)^{11/3}~\text{Gyr}$ \\
			\hline
			2 & $\displaystyle \frac{1.1 \times 10^{54}}{\sqrt{\beta}} \left(\frac{M}{\text{g}}\right)^{5}~\text{Gyr}$ \\
			\hline
		\end{tabular}
		\caption{Lifetime of 6D Near-Extremal Black Holes with ($p=1,2$) and without ($p=0$) Memory Burden Effect.}
		\label{tab:lifetimes}
	\end{table}
	These results demonstrate that the memory burden effect extends enormously the lifetime of near-extremal black holes, particularly for larger values of $p$.\footnote{The total lifetime of black hole is given by:  $\tau_{\text{total}} = \tau_{\text{semi}} + \tau_{\text{MB}}$. Because $S \gg 1$, $\tau_{\text{MB}} \gg \tau_{\text{semi}}$, making $\tau_{\text{MB}}$ the dominant contribution while formally accounting for the initial semiclassical regime.} For $p \geq 1$, even sub-gram black holes can survive to the present day, opening up new regions of parameter space for PBH dark matter in the two-dark-dimensions scenario.

    \paragraph{}
    \begin{figure}[h]
	\centering
	\includegraphics[width=0.8\textwidth]{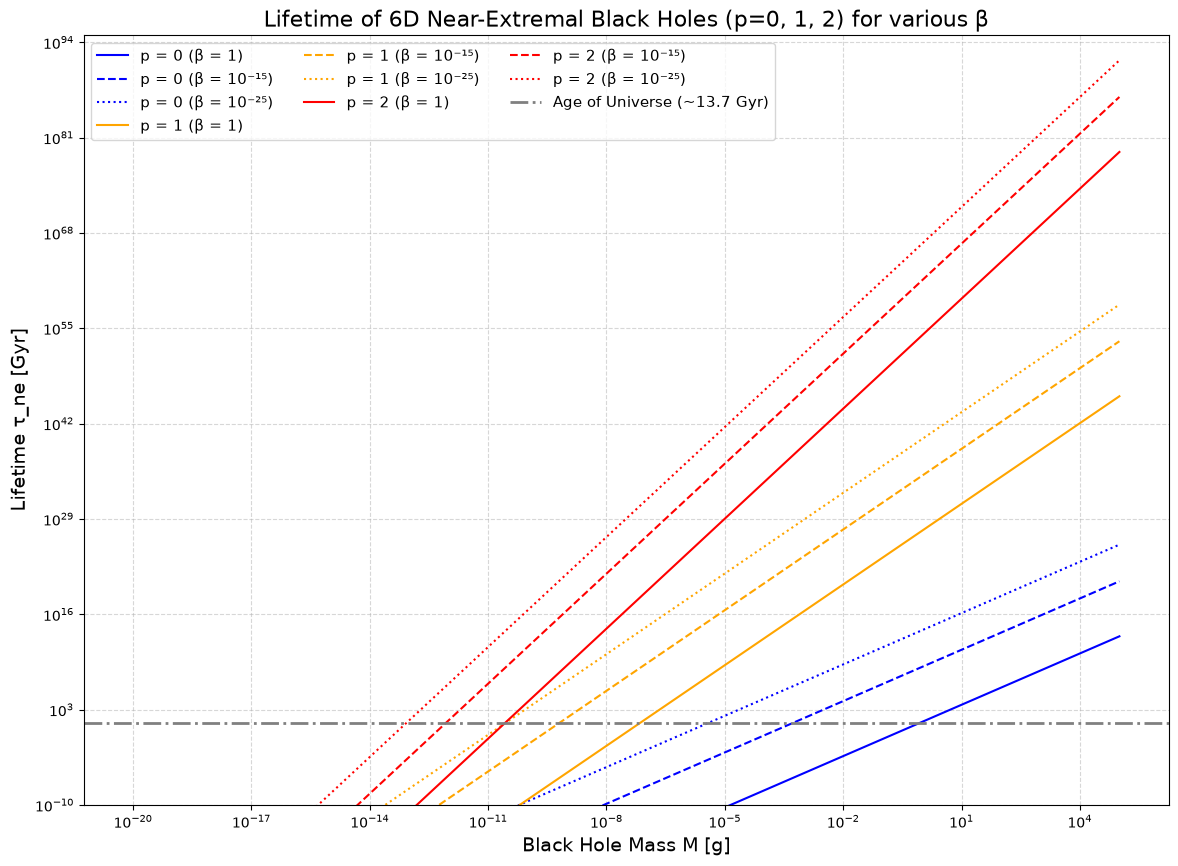}
	\caption{Lifetime $\tau_{ne}$ of 6D near-extremal primordial black holes as a function of their mass $M$. The blue curves represent the standard Hawking evaporation scenario ($p=0$), while the orange ($p=1$) and red ($p=2$) curves illustrate the suppressed decay rates due to the memory burden effect. The line styles denote varying deviations from extremality, parametrized by $\beta$: solid lines correspond to $\beta = 1$, dashed lines to highly extreme configurations ($\beta = 10^{-15}$), and dotted lines to ultra-extreme cases ($\beta = 10^{-25}$). The horizontal grey dash-dotted line marks the current age of the Universe ($\sim 13.7$ Gyr). As shown, both the memory burden effect and a strong extremality limit independently as well as cooperatively prolong the black hole lifetime, allowing even sub-gram mass black holes to survive to the present epoch and act as viable dark matter candidates. }
	\label{betas}
\end{figure}
Finally, it is important to see the influence of the parameter $\beta$ in the lifetime of near-extremal primordial black holes. In the Figure \ref{betas}  we can see the lifetime of 6D near-extremal primordial black holes as a function of their mass M for three different values of parameter $\beta$. The most critical cosmological implication of these findings is determined by the intersection of these curves with the horizontal grey dash-dotted line, which marks the current age of the Universe ($\sim 13.7$ Gyr). For a PBH to be a viable dark matter candidate today, its mass must lie to the right of this intersection. While standard Hawking evaporation requires relatively massive PBHs to evade complete evaporation, the inclusion of the memory burden effect—especially when coupled with ultra-extreme configurations—shifts this viability threshold deep into the sub-gram regime. Ultimately, the plot confirms that under the combined suppression of near-extremality and memory burden, microscopic, sub-gram black holes that would classically evaporate almost instantaneously can remain stable up to the present epoch, opening a vast and novel parameter space for dark matter in the two-dark-dimensions scenario.

    .
	
	\section{Higher-Dimensional Rotating Black Holes}

Our primary goal in this section is to determine whether 5D and 6D rotating primordial black holes (PBHs) can constitute the dark matter of the present universe. This requires a reliable estimate of their lifetimes, which depend on mass and the evaporation mechanism. We therefore present computational details towards the derivation of black hole lifetimes. Because the lifetime scales with mass, a natural mass bound emerges, delineating which PBHs can contribute to dark matter and whether they can account for all of it.
    
	\subsection{Lifetime of 5D Rotating Black Holes}
	
	In Ref.~\cite{Leontaris:2025piz}, we calculated the lifetime of 5D rotating black holes in the Dark Dimension scenario. In this section, we summarize the basic results of that work and next we extend them to the  six  dimensions. 
    
    The well-known relation for the mass decay rate is,
	\begin{equation}\label{eq:dMdt_general}
		\frac{dM}{dt} = -\frac{1}{2\pi r_H^2} \sum_{s,l,m} \int_0^\infty \frac{\tilde{\omega} \, \Gamma_{s,l,m}(\tilde{\omega})}{e^{(\tilde{\omega} - m\tilde{\Omega})/\tilde{T}} - (-1)^{2s}} \, d\tilde{\omega}
		= -\frac{C_M(a_*)}{r_H^2},
	\end{equation}
	where $r_H$ is the horizon radius, $\Gamma_{s,l,m}$ are the greybody factors, $\tilde{\Omega}$ is the angular velocity, and $C_M(a_*)$ is a dimensionless function of the dimensionless spin parameter $a_* = a/r_H$. For the 5D case, this yields the mass decay rate,
	\begin{equation}\label{eq:dmdt_5d}
		\frac{dM}{dt} \simeq -5.68 \times 10^{29} \, \frac{\text{GeV}^3}{M}.
	\end{equation}
	
	From this decay rate, one obtains the lifetime of the black hole,
	\begin{equation}\label{lifetime_5d_rot}
		\tau^{n=1}_{\text{MP}} = \tau^{\text{s.d.}}_{\text{BH}} + \tau^{\text{S.T.}}_{\text{BH}} \simeq 3.5 \times 10^{9}~\text{yr},
	\end{equation}
	where the two contributions correspond to the spin-down and Schwarzschild phases, respectively. Consequently, if 5D rotating black holes are to constitute all of the dark matter, the viable mass range is,
	\begin{equation}\label{massrange_5d_rot}
		10^{14} \lesssim M_{\text{BH}}/\text{g} \lesssim 10^{21}.
	\end{equation}
	
	When the memory burden effect is taken into account, the allowed parameter space expands further. The lifetimes for the cases $p=1$ and $p=2$ are summarized in Table~\ref{tab:MB_5d}. For comparison purposes, we have also included the $p=0$ case (no memory burden effect, just standard  Hawking radiation)  in this Table. 
	
	\begin{table}[htbp]
		\centering
		\begin{tabular}{|c|c|c|}
			\hline
			$p$ & $\tau(M)$ & Transition to MB phase \\
			\hline
			$0$ & $\displaystyle \frac{1}{6\pi^2} \frac{1}{C_M} \left(\frac{M}{M_*}\right)^2 \frac{1}{M_*} \approx 2.78 \times 10^{-15} \left(\frac{M}{\text{g}}\right)^2 \text{yr}$ & Hawking radiation \\
			\hline
			$1$ & $\displaystyle \frac{8}{63\sqrt{3}\pi^2} \frac{1}{C_M} \left(\frac{M}{M_*}\right)^{7/2} \frac{1}{M_*} \approx 5.15 \times 10^{7} \left(\frac{M}{\text{g}}\right)^{7/2} \text{yr}$ & Instantaneous MB effect \\
			\hline
			$2$ & $\displaystyle \frac{1}{5\pi^2} \left(\frac{2}{3}\right)^4 \frac{1}{C_M} \left(\frac{M}{M_*}\right)^{5} \frac{1}{M_*} \approx 1.17 \times 10^{26} \left(\frac{M}{\text{g}}\right)^5 \text{yr}$ & Semiclassical description, \\
			& &breaks down to MB phase \\
			\hline
		\end{tabular}
		\caption{Lifetime of a 5D Rotating PBH as a function of mass for three values of the exponent $p$ in the memory burden  effect ($p=0$ corresponds to standard Hawking radiation).}
		\label{tab:MB_5d}
	\end{table}
	
	\subsection{Lifetime of 6D Rotating Black Holes}
	
	The Hawking evaporation time characterizing the lifetime of a bulk PBH is estimated to be~\cite{TwoDarkDim},
	\begin{equation}\label{lifetime_general}
		\tau_{\text{BH}} \sim r_h S_{\text{BH}}.
	\end{equation}
	
	For the case $\mathfrak{n}=2$ ($D=6$), the event horizon radius for a rotating black hole is,
	\begin{equation}\label{horizon_6d_rot}
		r_h \sim \frac{1}{(1 + a_*^2)^{1/3}} \frac{1}{M_*} \left(\frac{M_{\text{BH}}}{M_*}\right)^{1/3},
	\end{equation}
	where $a_* = a/r_h$ is the dimensionless spin parameter. The 6D Newton constant is $G_6 = 1/(8\pi M_*^4)$, and the area of the unit 4-sphere is 
	$$A_4 = 2\pi^{5/2}/\Gamma(5/2) = 2\pi^{5/2}/(3\sqrt{\pi}/4) = 8\pi^2/3~.$$ The mass parameter is defined as,
	\begin{equation}
		\mu = \frac{16\pi G_6}{4A_4} M_{\text{BH}} = \frac{16\pi}{8\pi M_*^4} \cdot \frac{3}{8\pi^2} M_{\text{BH}} = \frac{3}{4\pi^2} \frac{M_{\text{BH}}}{M_*^4},
	\end{equation}
	and the rotational parameter $a = J/(2M_{\text{BH}})$, with $J$ the angular momentum.
	
	The horizon condition for a 6D rotating black hole is obtained from $g^{rr}=0$,
	\begin{equation}
		r^2 + a^2 - \frac{\mu}{r} = 0 \quad \Rightarrow \quad \mu = r_+(r_+^2 + a^2).
	\end{equation}
	The entropy of a 6D rotating black hole is:
	\begin{equation}\label{entropy_6d_rot}
		S_{\text{BH}} = \frac{2\pi r_h^2 (r_h^2 + a^2)}{3 G_6} = \frac{16\pi^2}{3} M_*^4 r_h^2 (r_h^2 + a^2).
	\end{equation}
	In terms of the dimensionless spin parameter $a_* = a/r_h$, this becomes,
	\begin{equation}\label{entropy_6d_rot_dimless}
		S_{\text{BH}} = \frac{16\pi^2}{3} M_*^4 r_h^4 (1 + a_*^2).
	\end{equation}
	Using Eq.~(\ref{horizon_6d_rot}) for $r_h$, we obtain,
	\begin{equation}\label{entropy_6d_rot_final}
		S_{\text{BH}} = \frac{16\pi^2}{3} \frac{1}{(1 + a_*^2)^{1/3}} \left(\frac{M_{\text{BH}}}{M_*}\right)^{4/3}.
	\end{equation}
	
	Finally, the lifetime of a 6D rotating black hole is given by,
	\begin{equation}\label{lifetime_6d_rot}
		\tau^{n=1}_{\text{MP}} = \tau^{\text{s.d.}}_{\text{BH}} + \tau^{\text{S.T.}}_{\text{BH}} \sim \left(1+\frac{2\pi}{3 (1 + a_*^2)^{2/3}}\right) \frac{1}{M_*} \left(\frac{M_{\text{BH}}}{M_*}\right)^{5/3}.
	\end{equation}
	
	\subsection{Memory Burden Effect for 6D Rotating Black Holes}
	
	We now suppose that the memory burden effect governs the evaporation process after the Page time~\cite{Page:1993wv}. The memory burden phase starts when the  rotation phase ends and therefore the remaining black hole has entered the Schwarzchild phase. From Ref.~\cite{Leontaris:2025piz} and the relation for the decay rate in $D$-dimensions we have,
	\begin{equation}
		-\frac{dM}{dt} = c_{\mathfrak{n}} \left(\frac{\mathfrak{n}+1}{4\pi}\right)^{\mathfrak{n}+2} T_H^2 \sim \left(\frac{\mathfrak{n}+1}{4\pi}\right)^{\mathfrak{n}+4} \frac{1}{r_s^2},
	\end{equation}
	where $c_{\mathfrak{n}}$ is a dimensionless constant that depends on the number of extra dimensions
	and the spectrum of emitted particles. For the case $\mathfrak{n}=2$, this yields,
	\begin{equation}
		-\frac{dM}{dt} \sim \left(\frac{3}{4\pi}\right)^6 \frac{1}{r_s^2}.
	\end{equation}
	
	Under the memory burden effect, the quantum mass decay rate is suppressed by a power $p$ of the entropy,
	\begin{equation}
		\frac{dM_{\text{MB}}}{dt} = \frac{1}{S^p} \frac{dM_{\text{SC}}}{dt},
	\end{equation}
	where $p$ is the memory burden exponent. For a 6D black hole, the entropy is found to be of the order
	\begin{equation}
		S_{\text{BH}} \sim \left(\frac{M_{\text{BH}}}{M_*}\right)^{4/3}.
	\end{equation}
	Then the relation for the lifetime under memory burden is given,
	\begin{equation}
		\tau_{MB}^p\sim \left(\frac{3}{4\pi}\right)^{-6}\left(\frac{4p+5}{3}\right)^{-1} M_{BH}^{\frac{4p+5}{3}}\left(\frac{1}{M_*}\right)^{\frac{4p+8}{3}} 
	\end{equation}
	For $p=1$, the lifetime under the memory burden effect is,
	\begin{equation}
		\tau_{\text{MB}}^{p=1} \sim \frac{1}{\gamma M_*} \left(\frac{M}{M_*}\right)^{3},
	\end{equation}
	where the numerical factor $\gamma$ incorporates various coefficients and is estimated to be $\gamma \approx 5.5\times10^{-4}$~\footnote{The parameters $\gamma$ and $\zeta$ constitute a product of multiplicative factors derived during the steps of the calculation of lifetimes in p=1 and p=2 respectively}. For $p=2$, we obtain,
	\begin{equation}
		\tau_{\text{MB}}^{p=2} \sim \frac{1}{\zeta M_*} \left(\frac{M}{M_*}\right)^{13/3},
	\end{equation}
	where the coefficient $\zeta$ takes the value $\zeta\approx 8\times10^{-4}$.
	
	Converting to units of mass in grams and time in gigayears, using $M_* = 10$ TeV $\sim 1.78 \times 10^{-32}$ g and $1~\text{Gyr} = 5.1 \times 10^{41}~\text{GeV}^{-1}$, we find,
	\begin{align}
		\tau_{\text{MB}}^{p=1} &\sim \frac{3.6}{\gamma} \times 10^{14} \left(\frac{M_{\text{BH}}}{\text{g}}\right)^{3}~\text{Gyr}, \\
		\tau_{\text{MB}}^{p=2} &\sim \frac{1.2}{\zeta} \times 10^{40} \left(\frac{M_{\text{BH}}}{\text{g}}\right)^{13/3}~\text{Gyr}.
	\end{align}
	
	For completeness, the $p=0$ case (standard Hawking radiation) gives,
	\begin{align}
	    \tau_{\text{MB}}^{p=0}& =\tau^{\text{s.d.}}_{\text{BH}} + \tau^{\text{S.T.}}_{\text{BH}}\sim \tau^{\text{s.d.}}_{\text{BH}} \nonumber \\  &\sim \left(\frac{2\pi}{3 (1 + a_*^2)^{2/3}}\right) \frac{1}{\xi M_*} \left(\frac{M_{\text{BH}}}{M_*}\right)^{5/3} \sim \frac{1.3}{\xi} \times 10^{-13} \left(\frac{M_{\text{BH}}}{\text{g}}\right)^{5/3}~\text{Gyr},
	\end{align}
	where we have taken $a_*\sim 1$ as a typical value and $\xi \approx3\times10^{-4}$. In the above calculation, we made the approximation that the black hole spends its entire life in the spin down phase.But for more concrete results you should calculate the lifetime of black hole spends in the spin down phase and in the final stage when the rotation stops.
	
	In Table~\ref{tab:MB_6d}, we collect the results for the lifetime of 6D Rotating Black Holes with memory burden effect, for the cases $p=1,2$. Again, for completeness we also include the case $p=0$ which corresponds to the standard Hawking radiation.
	
	\begin{table}[htbp]
		\centering
		\begin{tabular}{|c|c|}
			\hline
			\textbf{$p$} & \textbf{$\tau_{\text{MB}}$} \\
			\hline
			0 & $\displaystyle  \frac{1.3}{\xi} \times 10^{-13} \left(\frac{M_{\text{BH}}}{\text{g}}\right)^{5/3}~\text{Gyr}$ \\
			\hline
			1 & $\displaystyle \frac{3.6}{\gamma} \times 10^{14} \left(\frac{M_{\text{BH}}}{\text{g}}\right)^{3}~\text{Gyr}$ \\
			\hline
			2 & $\displaystyle \frac{1.2}{\zeta} \times 10^{40} \left(\frac{M_{\text{BH}}}{\text{g}}\right)^{13/3}~\text{Gyr}$ \\
			\hline
		\end{tabular}
		\caption{Lifetime of 6D Rotating Black Holes with Memory Burden Effect for $p=1,2$. The value $p=0$  corresponds to Hawking evaporation.}
		\label{tab:MB_6d}
	\end{table}
	
These findings show that, in the memory burden scenario, rotating six-dimensional black holes experience an enormous  prolongation of their lifetime compared to standard Hawking  evaporation.  Even for the smallest  $p=1,2$ exponents,  PBHs  with mass  as low as a few sub-grams can survive to the present day, significantly expanding the viable parameter space for PBH dark matter in the two-dark-dimensions scenario.

\begin{figure}[h]
	\centering
	\includegraphics[width=0.8\textwidth]{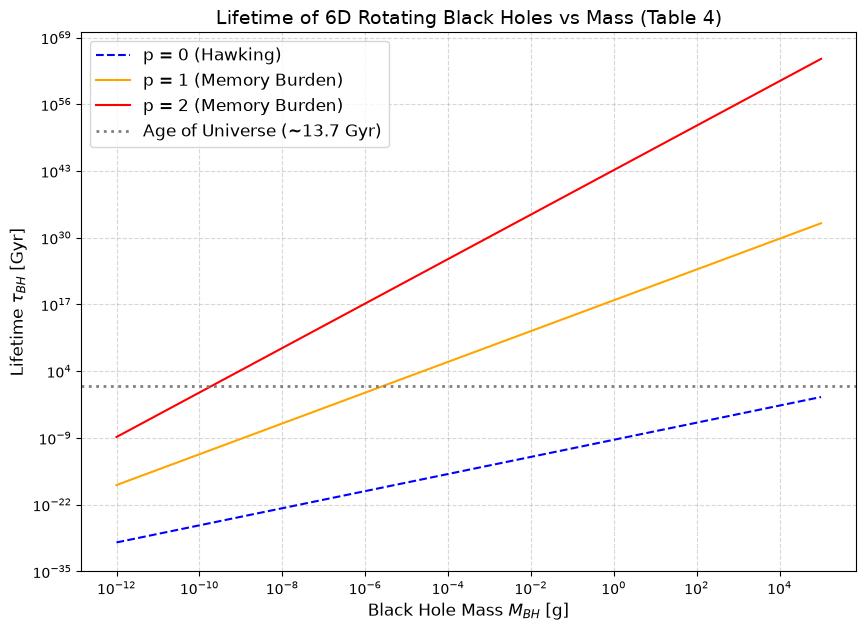}
	\caption{ illustrates the lifetime of 6D rotating primordial black holes across a broad mass spectrum ($10^{-12} \text{ g} \le M_{BH} \le 10^5 \text{ g}$) for $p=0$ (pure Hawking evaporation) as well as $p=1, 2$ (memory burden scenario). In the absence of memory burden ($p=0$), sub-gram rotating black holes evaporate almost instantaneously, with lifetimes falling well below $10^{-12} \text{ Gyr}$. However, once the memory burden phase sets in after the Page time, the quantum suppression by powers of entropy ($S^p$) dramatically alters the evaporation trajectory. For $p=1$, PBHs with masses as low as $M_{BH} \gtrsim 10^{-4.5} \text{ g}$ survive to the present day, while for $p=2$, even ultra-light PBHs down to $M_{BH} \gtrsim 10^{-9.5} \text{ g}$ remain stable over cosmological timescales. This visual comparison underlines how the memory burden effect opens up a vast new mass window for rotating PBH dark matter in two extra dark dimensions.}
	\label{PBHlifeB}
\end{figure}

	\section{Footprints of 6D Micro Black Holes at Future Colliders}
	
	 High energy collisions at future colliders offer a unique opportunity to investigate the existence of 6D
primordial  black holes.   This can be detected through distinctive traces resulting from the scattering of gravitons in the extra dimensions
	and the subsequent decay of these mini black holes into observable Standard Model particles.

	In the case of $\mathfrak{n}=2$ extra dimensions, the fundamental Planck scale is related to the four-dimensional Planck mass by,
	\begin{equation}\label{planck_relation}
		M_P^2 = R^{\mathfrak{n}} M_*^{\mathfrak{n}+2} \quad \Longrightarrow \quad M_P^2 = R^2 M_*^4 \quad \text{(for } \mathfrak{n}=2\text{)},
	\end{equation}
	where $R$ is the compactification radius of the extra dimensions, and the reduced Planck mass is $\overline{M}_P = M_P / \sqrt{8\pi} \approx 2.4 \times 10^{18}$ GeV.
	
	\noindent 
	The mass splitting $\Delta m$ between KK  modes is related to the radius $R$~\cite{ADD2}:
	\begin{equation}\label{mass_splitting}
		\Delta m \sim \frac{1}{R} = M_* \left( \frac{M_*}{\overline{M}_P} \right)^{\frac{2}{\mathfrak{n}}}.
	\end{equation}
	For $\mathfrak{n}=2$, this becomes,
	\begin{equation}\label{mass_splitting_n2}
		\Delta m \sim \frac{M_*^2}{\overline{M}_P}.
	\end{equation}
	With $M_* \approx 10$ TeV and $\overline{M}_P \approx 2.4 \times 10^{15}$ TeV, we obtain
	\begin{equation}
		\Delta m \approx \frac{(10 \text{ TeV})^2}{2.4 \times 10^{15} \text{ TeV}} \approx 4.2 \times 10^{-14} \text{ TeV} \approx 4.2 \times 10^{-2} \text{ eV}~,
	\end{equation}
	
	Remarkably, this value falls in the same range as the observed atmospheric neutrino mass scale $\sqrt{\Delta m^2_{\text{atm}}} \sim 0.05$ eV~\cite{King:2014nza}. This coincidence is not accidental within the Dark Dimension scenario: the KK gravitons or other bulk fields with masses of order $\Delta m$ could mix with active neutrinos via higher-dimensional operators, naturally generating the small observed neutrino masses. Indeed, in the original Dark Dimension proposal~\cite{DarkDim}, the lightest KK modes with masses $\sim 10^{-3}$ eV were suggested as potential dark radiation candidates. The  extension to $\mathfrak{n}=2 $ dark dimensions yields a slightly higher mass scale $\sim 4 \times 10^{-2}$ eV, which is precisely where the atmospheric neutrino mass difference resides. This provides a remarkable hint that the dark dimension may be responsible for both dark energy and the neutrino mass hierarchy.

	\subsection{Virtual Graviton Exchange}
	
	We now study the effect of a single virtual graviton exchange at tree level in scattering processes, following  Ref.~\cite{Giudice:1998ck}. For simplicity, we consider the case of pure $s$-channel exchange, though the $t$-channel and $u$-channel processes are analogous. The scattering amplitude for a graviton-mediated process in momentum space is given by:
	\begin{equation}\label{scat_ampl}
		\mathcal{A} = \frac{1}{\overline{M}_P^2} \sum_n \left( T_{\mu\nu} \frac{P^{\mu\nu\alpha\beta}}{s - m^2} T_{\alpha\beta} + \left(\frac{\kappa}{3}\right)^2 \frac{T_\mu^\nu T_\nu^\mu}{s - m^2} \right) = S(s) \, T,
	\end{equation}
	where
	\begin{equation}
		S(s) = \frac{1}{\overline{M}_P^2} \sum_n \frac{1}{s - m^2}, \qquad
		T = T_{\mu\nu} T^{\mu\nu} - \frac{1}{n+2} T^\mu_\mu T^\nu_\nu,
	\end{equation}
	with $T_{\mu\nu}$ the energy-momentum tensor, $k$ the transferred momentum, $\kappa = \sqrt{\frac{3(n-1)}{n+2}}$, and the graviton polarization sum:
	\begin{align}
		P_{\mu\nu\alpha\beta} &= \frac{1}{2} (\eta_{\mu\alpha}\eta_{\nu\beta} + \eta_{\mu\beta}\eta_{\nu\alpha} - \eta_{\mu\nu}\eta_{\alpha\beta}) \nonumber \\
		&\quad - \frac{1}{2m^2} (\eta_{\mu\alpha}k_\nu k_\beta + \eta_{\mu\beta}k_\nu k_\alpha + \eta_{\nu\alpha}k_\mu k_\beta + \eta_{\nu\beta}k_\mu k_\alpha) \nonumber \\
		&\quad + \frac{1}{6} \left(\eta_{\mu\nu} + \frac{2}{m^2}k_\mu k_\nu\right) \left(\eta_{\alpha\beta} + \frac{2}{m^2}k_\alpha k_\beta\right).
	\end{align}
	
	The two terms in Eq.~(\ref{scat_ampl}) correspond to the exchange of the graviton $G^{(n)}_{\mu\nu}$ and the scalar $H^{(n)}$, respectively. The sum $\sum_n$ runs over all KK modes and must be performed at the amplitude level. Because the interaction operator $T$ does not depend on which specific KK mode is involved, the total sum is independent of the specific particle scattering process under consideration. Converting the sum to an integral and evaluating it using dimensional regularization, we obtain for even $\mathfrak{n}$ ($\mathfrak{n} \geq 2$),
	\begin{equation}
		S(s) = -\frac{1}{M_*^{2+\mathfrak{n}}} \frac{S_{\mathfrak{n}-1}}{2} \left( \left(i\pi + \ln\left(\frac{s}{\mu^2}\right)\right) s^{(\mathfrak{n}-2)/2} + \sum_{k}^{(\mathfrak{n}-2)/2} c_k \Lambda^{\mathfrak{n}-2k} s^{k-1} \right),
	\end{equation}
	and for odd $n$ 
	\begin{equation}
		S(s) = -\frac{1}{M_*^{2+\mathfrak{n}}} \frac{S_{\mathfrak{n}-1}}{2} \left( i\pi \sqrt{s} \, s^{(\mathfrak{n}-3)/2} + \sum_{k}^{(\mathfrak{\mathfrak{n}}-1)/2} c_k \Lambda^{\mathfrak{n}-2k} s^{k-1} \right),
	\end{equation}
	where $S_{\mathfrak{n}-1} = 2\pi^{(\mathfrak{n}-1)/2}/\Gamma((\mathfrak{n}-1)/2)$ is the area of the $(\mathfrak{n}-1)$-sphere, $\Lambda$ is a UV cutoff, and $c_k$ are numerical coefficients.
	
	For our case of interest, $\mathfrak{n}=2$, we obtain,
	\begin{equation}\label{S_n2}
		S(s) = -\frac{\pi}{M_*^4} \left( i\pi + \ln\left(\frac{s}{\mu^2}\right) \right).
	\end{equation}
	
	\subsection{Differential Cross Sections}
	
	The differential cross section for inclusive graviton production for $\mathfrak{n}=2$ is expressed as:
	\begin{equation}
		\frac{d^2\sigma}{dt\,dm} = 2\pi \frac{\overline{M}_P^2}{M_*^4} \, m \, \frac{d\sigma}{dt},
	\end{equation}
	where $m$ is the invariant mass of the KK graviton.
	
	For the process $f\bar{f} \to \gamma G$, we have,
	\begin{equation}
		\frac{d\sigma}{dt} = \frac{\alpha Q_f^2}{16 N_f} \frac{1}{s \overline{M}_P^2} F_1(s, m^2, t).
	\end{equation}
	
	For parton-level QCD processes, the differential cross sections are:
	\begin{align}
		\frac{d\sigma}{dt}(q\bar{q} \to gG) &= \frac{\alpha_s}{36} \frac{1}{s \overline{M}_P^2} F_1, \\
		\frac{d\sigma}{dt}(qg \to qG) &= \frac{\alpha_s}{96} \frac{1}{s \overline{M}_P^2} F_2, \\
		\frac{d\sigma}{dt}(gg \to gG) &= \frac{3\alpha_s}{16} \frac{1}{s \overline{M}_P^2} F_3, \label{cross_grav}
	\end{align}
	where the functions $F_1$, $F_2$, and $F_3$ can be found in Appendix A of Ref.~\cite{Giudice:1998ck}.
	
	\subsection{Measuring the Multiplicity}
	
	In the context of particle physics, \textit{multiplicity} refers to the total number of individual particles produced in high-energy processes such as particle collisions or the decay of a mini black hole. When a black hole is created in a collider, it evaporates almost instantly, exploding into a wide variety of secondary particles. The multiplicity is the count of those particles that reach the detector.
	
	Multiplicity is one of the most powerful discriminators between black hole events and Standard Model backgrounds. A typical Standard Model event produces a small number of high-energy particles, whereas a black hole event yields a high multiplicity of particles. A single mini black hole might decay into $10$ to $25$ distinct particles.
	
	For $\mathfrak{n}=2$ extra dimensions and $M_* = 10$ TeV, the Hawking temperature for a Schwarzschild micro black hole is given by,
	\begin{equation}
		T_H = \frac{\mathfrak{n}+1}{4\pi r_h} \quad \Longrightarrow \quad T_H \approx \frac{3}{4\pi} \frac{M_*^{4/3}}{M_{\text{BH}}^{1/3}}.
	\end{equation}
	
	The average multiplicity $\langle N \rangle$ of decay products is,
	\begin{equation}\label{multiplicity}
		\langle N \rangle = \frac{M_{\text{BH}}}{2 T_H} \approx \frac{2\pi}{3} \left( \frac{M_{\text{BH}}}{M_*} \right)^{4/3}.
	\end{equation}
	Assuming the black hole mass remains constant during the evaporation phase, $\langle N \rangle$ depends only on $T_H$.
	
	In the two-dark-dimensions scenario with $M_* = 10$ TeV, we calculate the multiplicity for various black hole masses
	\begin{itemize}
		\item $M_{\text{BH}} = 12$ TeV: $\displaystyle \langle N \rangle \approx \frac{2\pi}{3} \left(\frac{12}{10}\right)^{4/3} \approx 3$
		\item $M_{\text{BH}} = 20$ TeV: $\displaystyle \langle N \rangle \approx \frac{2\pi}{3} \left(\frac{20}{10}\right)^{4/3} \approx 5$
		\item $M_{\text{BH}} = 30$ TeV: $\displaystyle \langle N \rangle \approx \frac{2\pi}{3} \left(\frac{30}{10}\right)^{4/3} \approx 9$
	\end{itemize}
	
	The most interesting case is $M_{\text{BH}} = 100$ TeV, which corresponds to the energy scale of the proposed Future Circular Collider (FCC):
	\begin{equation}
		\langle N \rangle \approx \frac{2\pi}{3} \left(\frac{100}{10}\right)^{4/3} \approx 45.
	\end{equation}
	
	Detecting a high-multiplicity event with $\langle N \rangle \sim 45$ and a thermal distribution would be a distinctive hallmark of a micro black hole at the FCC. Moreover, by measuring the multiplicity and mass of a large number of black hole evaporation events with high precision, one can extract the fundamental Planck scale $M_*$ from Eq.~(\ref{multiplicity}).
	
	\subsection{Measuring the Temperature}
	
	The energy threshold for black hole production at the LHC is given by the condition $\lambda_C < r_h$, where $\lambda_C$ is the Compton wavelength
	\begin{equation}
		\lambda_C = \frac{4\pi}{E} < \frac{1}{\sqrt{\pi} M_*} \left(\frac{E}{M_*}\right)^{\frac{1}{\mathfrak{n}+1}} \left( \frac{8\,\Gamma\!\left(\frac{\mathfrak{n}+3}{2}\right)}{\mathfrak{n}+2} \right)^{\frac{1}{\mathfrak{n}+1}},
	\end{equation}
	with $E$ being the total energy of the colliding partons that form the black hole.
	
	The production cross-section in the center-of-mass energy of the collision is proportional to the horizon area
	\begin{equation}
		\sigma_{\text{prod}} \propto \pi r_h^2 \sim \frac{1}{M_*^2} \left( \frac{E}{M_*} \right)^{\frac{2}{\mathfrak{n}+1}}.
	\end{equation}
	This relation reveals an enhancement of the cross-section with $E$ that is not present in any Standard Model or SUSY process, providing a distinct  signature.
	
	To extract useful information about  the scenario with  two dark dimensions, we must accurately measure  the mass and the temperature of the produced mini black holes. In a high-energy collision of composite particles, it is known which pair of partons led to the creation of the black hole, nor what its total energy was. Therefore, the black hole mass must   be reconstructed, through the measurement of the energy of the particles appearing in the final state after evaporation.
	The temperature of the black hole can be determined by fitting the observed Hawking radiation spectrum. Particularly clean spectra can be obtained from events containing only photons and electrons in the final state, which offer very low backgrounds and excellent energy resolution even at high energies.

	From the Hawking temperature relation, we can write
	\begin{equation}
		\log(T_H) = -\frac{1}{\mathfrak{n}+1} \log(M_{\text{BH}}) + \text{constant}.
	\end{equation}
	For $\mathfrak{n}=2$, this becomes
	\begin{equation}
		\log(T_H) = -\frac{1}{3} \log(M_{\text{BH}}) + C,\label{LogLog}
	\end{equation}
where  the constant $C$ is given by 
$$C=\log 3 - \log\!\left( M_*^{4/3} \left( \frac{8\Gamma(5/2)}{5} \right)^{1/3} \right)~.$$
	Thus, by virtue of~(\ref{LogLog}) the number of extra dimensions, $\mathfrak{n}$, can be extracted by determining the slope of the straight-line fit relating $\log(M_{\text{BH}})$ and $\log(T_H)$.

	\section{Summary of Results and Discussion} 
	In this work, we investigated six-dimensional  primordial black holes within a model featuring two micron-sized dark extra dimensions,
	where the fundamental Planck scale is lowered down to $M_* \sim 10$ TeV. We considered charged and rotating black holes
	and explored their potential to account for the dark matter of the universe,  under the standard Hawking evaporation as  well as  under the memory burden scenario. Our results reveal a striking contrast between standard semiclassical Hawking evaporation and the memory burden scenario.

	Under standard Hawking evaporation, static six-dimensional PBHs with masses $  M \gtrsim 10^8  $ g survive to the present day, with lifetimes scaling as $  \tau \sim 13.7 \left( M / 10^8 \, \text{g} \right)^{5/3}  $ Gyr. Near-extremal charged configurations, whose temperature is suppressed by a factor $  \sqrt{\beta / S}  $, further extend the viable mass range down to $  M \gtrsim \ \beta^{3/14}  $ g, where $S$ is the entropy and $\beta$ a small parameter measuring deviations form the extremality condition. This opens a significantly broader parameter space for PBH dark matter compared to their four-dimensional counterparts.
	When the memory burden effect is included, the evaporation rate is additionally suppressed by a factor inversely proportional to the black hole entropy. This enormous suppression fundamentally changes the picture, allowing even sub-gram PBHs to remain stable up to the present epoch, masses that would have evaporated almost instantaneously under classical Hawking radiation. In   Table~\ref{HandMBlifetimes} we summarize our results
	for the lifetime  and viable mass ranges of PBHs for the two scenarios discussed above.
	\begin{table}[htbp]
		\centering
		\begin{tabular}{l|l|l}
			\textbf{Scenarios:} & \textbf{Lifetime scaling as} & \textbf{Viable Mass } \\
			{ \bf Hawking $\&$ memory burden} &{ \bf function of PBH mass} & \textbf{range for DM} \\
			\hline
			Hawking ($p=0$), static/rotating & $\tau \sim  \frac{1.3}{\xi} \times 10^{-13} \left(\frac{M_{\text{BH}}}{\text{g}}\right)^{5/3}~\text{Gyr}$ & $M > 10^8$ g \\
			Hawking ($p=0$), near-extremal & $\tau \sim \frac{21.3}{\sqrt{\beta}}\,\left(\frac{M}{\,\text{g}}\right)^{7/3}  $ Gyr & $M > \beta^{3/14}$ g \\
			Memory Burden ($p=1$) near-extremal & $\tau \sim \frac{2.5}{\sqrt{\beta}}\,10^{27}\,\left(\frac{M}{\text{g}}\right)^{11/3}$ Gyr & $M > 10^{-7.5}$ g \\
			Memory Burden ($p=2$) near-extremal & $\tau \sim \frac{1.1}{\sqrt{\beta}}\,10^{54}\,\left(\frac{M}{\text{g}}\right)^{5}$ Gyr &$ M>^{-10.5}$ g\\
			Memory Burden ($p=1$) static/rotating & $\tau \sim \frac{3.6}{\gamma}\,10^{14}\,\left(\frac{M}{\text{g}}\right)^{3}$ Gyr & $M > 10^{-4.5}$ g \\
			Memory Burden ($p=2$) static/rotating & $\tau \sim \frac{1.2}{\zeta}\,10^{40}\,\left(\frac{M}{\text{g}}\right)^{13/3}$ Gyr &$ M>10^{-9.5}$ g\\
			
		\end{tabular}
		\caption{Summary of 6D PBH lifetimes and viable dark matter mass ranges under standard Hawking evaporation $(p=0)$ and memory burden effect for $p=1$, and $ p=2$.}
		\label{HandMBlifetimes}
	\end{table}

	Our scenario is falsifiable through direct experimental tests at future colliders. For a fundamental Planck scale $  M_* \sim 10  $ TeV, micro black holes could be produced at future facilities such as the Future Circular Collider. According to our analysis, these black holes would evaporate via Hawking radiation, yielding high-multiplicity final states with characteristic thermal spectra. For a black hole mass $  M_{\rm BH} = 100  $ TeV, the average particle multiplicity reaches $  \langle N \rangle \sim 45  $, significantly exceeding typical Standard Model backgrounds. The observation of such events would constitute a clean signature of
	evaporation of micro black holes, enabling a direct measurement of both $  M_*  $ and the number of extra dimensions $  \mathfrak{n}  $ through the temperature-mass relation
	$$\log T_H = -\frac{1}{ \mathfrak{n}+1} \log M_{\rm BH} + \text{constant}.$$
	
	In addition, the  PBH mass ranges   determined  in this work can be checked for consistency with current observational constraints. Thus, PBHs in the $10^8$–$10^{21}$ g range evade microlensing bounds due to their small size and the gamma-ray background from Hawking evaporation is suppressed by virtue of the suppression of the evaporation rate in our 6D scenario. For memory burden scenarios, the suppression is even more pronounced, rendering these PBHs effectively invisible to current searches while remaining viable dark matter candidates. Future experiments, including improved gamma-ray telescopes and gravitational wave observatories, may probe these scenarios further.
	
	Yet, the above scenario  is not without challenges and caveats. The most  notable  issues include the following.
	
	In hadron colliders such as the LHC, the black hole mass $  M_{\rm BH}  $ cannot be observed directly and must be reconstructed from visible decay products, including missing transverse energy from neutrinos or stable KK gravitons. The democratic nature of Hawking radiation, finite detector acceptance, and limited energy resolution, especially for hadronic jets, introduce substantial uncertainties in the reconstruction. Advanced machine learning and kinematic fitting can mitigate these effects, but irreducible missing energy imposes fundamental limitations.
	
	Another obstacle is that as the black hole loses mass, its Hawking temperature $  T_H  $ rises slowly in six dimensions, according to the relation $  T_H \sim M_{\rm BH}^{-1/3}  $. The emitted spectrum is a superposition of thermal spectra with evolving temperature, complicating the extraction of the initial $  T_H  $. This can be mitigated by selecting events with very high initial masses, where the fractional mass loss during the observable phase is small, approximating a fixed-temperature source. There is also the issue where  secondary particles not originating directly from the evaporating black hole may obscure the spectrum.  Addressing these challenges will be crucial for a successful experimental identification of micro black holes and the determination of the underlying higher-dimensional parameters. A multi-dimensional approach combining improved detector resolution and simulation techniques will be required.
	
	In conclusion, motivated by the Dark Dimension scenario with two micron-sized extra dimensions \cite{TwoDarkDim}, which naturally addresses both the electroweak-Planck hierarchy and the cosmological constant problem, we extended previous analyses to investigate the evaporation physics and viable mass range of six-dimensional primordial black holes. We find that these PBHs can have lifetimes comparable to the age of the universe, allowing them to serve as viable dark matter candidates.
	Interestingly, for $   \mathfrak{n}=2  $ extra dimensions, the mass splitting $  \Delta m  $ between Kaluza–Klein modes is of the order of the observed atmospheric neutrino mass scale, $  \sqrt{\Delta m^2_{\rm atm}} \sim 0.05  $ eV.
\vfill

\noindent 
    {\it Acknowlegments:} GKL would like to thank the staff of the Department of Nuclear and Particle Physics,
Faculty of Physics, at the National and Kapodistrian University of Athens for kind hospitality during various stages of this work.

\end{document}